\documentclass[pra,superscriptaddress,twocolumn,nofootinbib]{revtex4-2}

\usepackage{enumerate}
\usepackage{amsfonts,amssymb,amsmath}
\usepackage[]{graphics,graphicx,epsfig}
\usepackage{amsthm}
\usepackage{graphicx}
\usepackage{dcolumn}
\usepackage{natbib}
\usepackage{color}
\usepackage{multirow}
\usepackage{ulem}
\usepackage{bm}
\usepackage{url}
\usepackage{float}
\usepackage{ragged2e}
\usepackage{mathbbol}

\usepackage[colorlinks = true, citecolor=green, linkcolor=blue, urlcolor=blue]{hyperref}
\newcommand*{\fullref}[1]{\hyperref[{#1}]{\autoref*{#1} \nameref*{#1}}}

\newcommand{\norm}[1]{\left\lVert#1\right\rVert}

\newcommand{\tr}[1]{\text{Tr}\left[#1\right]}
\newcommand{\one}{\mathbb{1}}

\usepackage{orcidlink}

\begin{document}


\title{Efficient measure of information backflow with a quasistochastic process}

\author{Kelvin Onggadinata\orcidlink{0000-0003-2328-8329}}
\email{kelvin.onggadinata@ntu.edu.sg}
\affiliation{School of Physical and Mathematical Sciences, Nanyang Technological University, 21 Nanyang Link, Singapore 637371, Singapore}

\author{Teck Seng Koh\orcidlink{0000-0002-4030-8232}}
\email{kohteckseng@ntu.edu.sg}
\affiliation{School of Physical and Mathematical Sciences, Nanyang Technological University, 21 Nanyang Link, Singapore 637371, Singapore}

\date{\today}


\begin{abstract}
Characterization and quantification of non-Markovian dynamics in open quantum systems are topical issues in the rapidly developing field of quantum computation and quantum communication. A standard approach based on the notion of information backflow detects the flow of information from the environment back to the system. Numerous measures of information backflow have been proposed using different definitions of distinguishability between pairs of quantum states. These measures, however, necessitate optimization over the state space, which can be analytically challenging or numerically demanding. Here we propose an alternative witness and measure of information backflow that is explicitly state independent by utilizing the concept of quasiprobability representation and recent advances in the theory of majorization for quasiprobabilities. We illustrate its use over several paradigmatic examples, demonstrating consistent Markovian conditions with known results and also reported necessary and sufficient conditions for the qutrit system in a random unitary channel. The paper concludes with a discussion of the foundational implications of quantum dynamical evolution. 
\end{abstract}

\maketitle


\section{Introduction}

In recent years, the study of open quantum systems beyond the standard Markovian approximation has gained considerable interest. In Markovian dynamics, information in a system dissipates continuously to the environment. Interestingly, it has been observed that non-Markovian dynamics exhibit a memory effect such that information can flow back to the system in a revival of the quantum features \cite{breuer2016colloquium, li2018concepts, chruscinski2022dynamical}. This phenomenon has become a subject of interest in the study of decoherence \cite{zurek2003decoherence, cai2018non}, quantum thermodynamics \cite{landi2021irreversible}, and quantum technologies in general \cite{li2020non}. Thus, an efficient tool to identify and quantify non-Markovianity features is highly desirable. 

There have been many proposals for a witness of non-Markovianity with the central object being the quantum dynamical map $\Lambda_t$, $t\geq 0$, which encapsulates the evolution of the open system. In the standard Hilbert space formalism, the maps are described as completely positive and trace-preserving (CPTP) maps that can inform the evolution of a quantum state at any time $t$ via $\rho(t) = \Lambda_t[\rho(0)]$. To track information backflow, one is only required to use some notion of distinguishability between two initial states and track its evolution over time. The most natural choice is the trace distance between two states defined as
\begin{equation}
    D(\rho_1, \rho_2) = \frac{1}{2}\norm{\rho_1 - \rho_2}\, ,
\end{equation}
where $\norm{A} = \text{Tr}|A|$ is the trace norm of $A$. Due to its contractive property, it is then easy to see that
\begin{equation}
    D(\Lambda_t[\rho_1], \Lambda_t[\rho_2]) \leq D(\rho_1, \rho_2)\, ,
\end{equation}
where the left-hand side is a monotonically decreasing function for all time $t\geq 0$. When monotonicity holds at all times, we say that $\Lambda_t$ is Markovian. In contrast, there exists quantum dynamics that breaks the monotonicity for some time interval, which results in a temporary revival of quantum information into the system. In such cases, we say that $\Lambda_t$ is non-Markovian. Therefore, following \cite{breuer2009measure}, the Breuer-Laine-Piilo (BLP) measure defined as 
\begin{equation}\label{eq: blp measure}
    \mathcal{N}_{\text{BLP}} = \max_{\rho_1, \rho_2}\int_{\sigma > 0}dt\, \sigma(\rho_1,\rho_2; t)\, ,
\end{equation}
with $\sigma(\rho_1, \rho_2; t) \equiv \frac{d}{dt}D(\Lambda_t[\rho_1], \Lambda_t[\rho_2])$, quantifies the degree of memory effects. The integral is taken only whenever the information flow $\sigma(\rho_1,\rho_2;t)$ is positive, an indication of violating Markovianity. As non-Markovianity is a feature solely due to the dynamics, maximization over all pair states is taken to remove the state dependence.

The above is not the only existing measure of information backflow. Many other proposals utilizing alternative or more general definitions of distances, such as the Helstrom matrix \cite{chruscinski2011measures,wissmann2015generalized}, and also entropic formulation \cite{luo2012quantifying, megier2021entropic} have been carried out. To our knowledge, these measures suffer from exact computation as optimizing over state space can be challenging, especially for higher-dimensional systems. Even though progress has been made to simplify the optimization task \cite{liu2014locality}, it would be ideal to completely characterize and measure non-Markovianity without this complication. It is also noteworthy that there exists another notion of non-Markovianity based on divisibility \cite{rivas2010entanglement, chruscinski2018divisibility}. P- or CP-divisibility of a dynamical map concerns whether the intermediate map is still positive or completely positive, respectively, which is then associated with being Markovian. The hierarchical relation between divisibility and information backflow has been studied extensively, with the former constituting a stricter criterion for non-Markovianity. Unlike the distinguishability-based measure, measures of non-Markovianity based on indivisibility are free of optimization. Hence, in this work, we focus on information backflow as our notion of non-Markovianity and find an efficient measure for it.

In this paper, we propose a novel measure of non-Markovianity, in particular, information backflow, that requires no optimization component using the framework of a quasiprobability representation (QPR) \cite{ferrie2009framed, ferrie2011quasi}. The QPR has been gaining popularity as it provides an alternative mathematical formalism of quantum theories that is reminiscent of probability theory but with the appearance of negative values, widely attributed as the signature of nonclassicality \cite{spekkens2008negativity, veitch2012negative, howard2014contextuality, tan2020negativity, onggadinata2023simulations, onggadinata2024negativity}. As such, it has been particularly useful in the study of classical versus quantum theory, identifying resources for quantum advantage, and in general providing new insights about quantum theories. Following the latter point, we will further comment on the implication of our non-Markovianity witness from a quantum foundational perspective. We note that previous work \cite{richter2022witnessing} had considered the use of frame representation to develop a new measure. However, similar to the previously mentioned measures, it also involves maximizing over all possible pairs of initial states. Crucially, that work focused on studying non-Markovianity in continuous-variable systems, but here we focus on the discrete case. 


\section{Quasiprobability representation}

In the QPR, elements of quantum theories, such as states and unitaries, are described by vectors or matrices consisting of real values that typically obey a normalization constraint. To translate quantum objects defined in the standard Hilbert space formalism into its QPR, the mapping is achieved by the so-called frame and its dual frame \cite{ferrie2009framed, ferrie2011quasi}. Hence, it is also colloquially known as frame representation. In the following, we will introduce the mathematical formalism of QPR.

For a system in Hilbert space $\mathcal{H}^d$, a frame $\{F_j\}$ is defined as a set of operators that span the Hermitian space, that is, it forms in general an overcomplete basis for the space of all Hermitian operators acting on $\mathcal{H}^d$. For a given frame, the dual frame $\{G_j\}$ is also a complete basis of a Hermitian space that satisfies $A = \sum_{j}\tr{AF_j}G_j$ for all Hermitian operator $A$. Note that the dual frame is generally not unique except for the case when the frame is minimal, i.e., $|\{F_j\}| = d^2$. In this case, the frames obey the relation $\tr{F_jG_k} = \delta_{jk}$. The QPR of a state $\rho$, channel $\mathcal{E}$, and measurement effect $M$ is then obtained via
\begin{eqnarray}
\rho \longrightarrow q^{\rho}_j &=&  \tr{\rho F_j} \, ,\\
\mathcal{E} \longrightarrow S^{\mathcal{E}}_{jk} & = & \tr{F_j\mathcal{E}[G_k]}\, , \\
M \longrightarrow v^{M}_{j} &=& \tr{M G_j}\, .
\end{eqnarray}
Note that the above always return real values and they correspond to the entries of the column vector $\mathbf{q}^\rho = [q^\rho_{j}]$, square matrix $S^{\mathcal{E}} = [S^{\mathcal{E}}_{jk}]$ (with row index $j$ and column index $k$), and row vector $\mathbf{v}^M = [v^M_{j}]$. From here, the Born rule probability of measuring a state undergoing an evolution beforehand is simply calculated following vector-matrix operation:
\begin{eqnarray}
P(M|\rho, \mathcal{E}) = \tr{M\mathcal{E}[\rho]} &=& \mathbf{v}^M S^{\mathcal{E}}\mathbf{q}^\rho \nonumber \\
& = &  \sum_{jk} v^M_j S^{\mathcal{E}}_{jk} q^\rho_k \, .
\end{eqnarray}

Here, we consider a normalization condition for the frame operators, $\sum_j F_j = \one$. This implies that $\sum_{j} q^\rho_j = 1$, so $\mathbf{q}^\rho$ can be interpreted as a quasiprobability distribution.  Moreover, we also impose that $\tr{G_j} = 1\, \forall\, j$ to ensure that summing up the effects in a positive-operator-valued measure (POVM) returns a unit measurement $\sum_{k} v^{M_k}_j = 1\, \forall \, j$, where $\{M_k\,|\,M_k\geq 0, \sum_{k}M_k = \one\}$ is a POVM. As a result, $v^{M_k}_j$ can be thought of as a conditional quasiprobability distribution. Finally, based on the above condition, we also find that $\sum_{j} S^{\mathcal{E}}_{jk} = 1 \, \forall \, k$, which means that it is now a quasistochastic matrix. If the channel is unital, i.e., $\mathcal{E}[\one] = \one$, then $S^\mathcal{E}$ is quasi-bistochastic,\footnote{In some literature, ``bistochastic" is also referred as ``doubly stochastic."} $\sum_{j} S^{\mathcal{E}}_{jk} = \sum_k S^{\mathcal{E}}_{jk} = 1$. As can be seen, the elements of quantum theories have been described in a language that is similar to classical theories and hence is natural to use for the study of classical-quantum distinction. 


\section{Majorization of quasiprobability distributions}

To derive the main results of this paper, we make use of the observations made in \cite{koukoulekidis2022constraints}. There the authors extended the theory of majorization for quasiprobabilities and showed that a certain family of entropies is well defined and meaningful even for quasiprobabilities. Of particular interest, Theorem 9 of \cite{koukoulekidis2022constraints} states the following: ``If $\alpha = \frac{2a}{2b-1}$ for positive integers $a,b$ with $a\geq b$, then $H_\alpha(\mathbf{q})$ is well defined on the set of quasidistributions, and if $\mathbf{q} \succ \mathbf{q}'$ for two quasidistributions $\mathbf{q},\mathbf{q}'$ then $H_\alpha(\mathbf{q}) \leq H_\alpha(\mathbf{q}')$." Here $H_\alpha(\mathbf{q})$ is the R{\'e}nyi-$\alpha$ entropy \cite{renyi1961measures} defined as 
\begin{equation}
    H_\alpha(\mathbf{q}) = \frac{1}{1-\alpha}\log_2\left[\sum_{k}q_k^\alpha\right]
\end{equation}
and $\mathbf{q} \succ \mathbf{q}'$ denotes that $\mathbf{q}$ majorizes $\mathbf{q}'$ in the preorder sense such that there exists a bistochastic map $A$ that satisfies the relation $A\mathbf{q} = \mathbf{q}'$.

The result above establishes the Schur-concavity property and monotonicity of R{\'e}nyi-$\alpha$ entropy on the set of quasiprobability distributions under a subset of $\alpha$ values. Moreover, the choice of $\alpha$ also ensures that the function is real-valued and satisfies many of the properties that R{\'e}nyi originally invoked \cite{brandenburger2024signed}.


\section{Main results}

In the following, we take advantage of the result above to derive a new witness and measure of non-Markovianity. We make the choice of $\alpha =2$ for the R{\'e}nyi-$\alpha$ entropy. This form is also known as collision entropy, and its usage has been explored previously as the entropy of the Wigner function \cite{manfredi2000entropy}, in relation to the uncertainty principle \cite{bosyk2012collision}, and in axiomatization of quantum theories \cite{brukner2009information,brandenburger2022renyi,onggadinata2023qubits}. 

Considering a state transforming over time given by $\mathbf{q}(t) = S^{\Lambda_t}\mathbf{q}(0)$, where we define $\mathbf{q} \equiv \mathbf{q}^\rho$ for brevity, it follows from the above that
\begin{equation}\label{eq: basic monotonicity}
    H_2(\mathbf{q}(t)) \geq H_2(\mathbf{q}(0))
\end{equation}
if $S^{\Lambda_t}$ is bistochastic for all time $t\geq 0$. One can think of the R{\'e}nyi-2 entropy of the quantum state as measuring the disorder of the quantum state in the usual manner. States with high disorder (low information content) will have higher entropy than states with low disorder (high information content). Following this interpretation, Markovian dynamics will result in a monotonic loss of information to the environment and violation of the monotonicity condition indicates memory effects (information backflow). Our formulation here is distinct from the usual approach based on monotonicity of the trace distance and is more akin to the entropic-based formulation in \cite{luo2012quantifying}.

In the current form, one can define a measure of non-Markovianity but it would still entail a maximization over state. Surprisingly, we show that it is possible to further simplify the above relation and remove the dependence on state. Note that the R{\'e}nyi-2 entropy can be expressed more compactly as
\begin{equation}
    H_2(\mathbf{q}) = -\log\left[\sum_{k}q_k^2\right] = -\log\mathbf{q}^T\mathbf{q}\, ,
\end{equation}
where $\mathbf{q}^T$ is the transpose of $\mathbf{q}$ into a row vector. Therefore, Eq. \eqref{eq: basic monotonicity} can be written as $-\log[(S^{\Lambda_t}\mathbf{q})^T(S^{\Lambda_t}\mathbf{q})] = -\log[\mathbf{q}^T (S^{\Lambda_t})^T S^{\Lambda_t}\mathbf{q}]  \geq  -\log \mathbf{q}^T\mathbf{q}$ and can be simplified neatly as 
\begin{equation}\label{eq: non-markovianity witness}
 (S^{\Lambda_t})^TS^{\Lambda_t} \leq  \one   \, .
\end{equation}
Under Markovian condition, the quasi-stochastic map of a channel $\Lambda_t$ must then obey this monotonic relation. More explicitly, the eigenvalues of $(S^{\Lambda_t})^TS^{\Lambda_t}$ must be a monotonically decreasing function over time, and the violation of the monotonicity signifies non-Markovianity. As such, we define our measure of non-Markovianity as
\begin{eqnarray}\label{eq: non-markovianity measure}
\mathcal{N} & = &  \int_{\zeta > 0} dt\, \zeta(t) \nonumber \\
\text{with}\quad \zeta(t) &=& \frac{d}{dt}\norm{(S^{\Lambda_t})^TS^{\Lambda_t}}\, .
\end{eqnarray}
The measure defined here is similar in spirit to previously proposed measures such that they sum up over every instance when non-Markovianity is observed based on the respective witness but will yield different values in general. This will be more evident in the examples below. Finally, we remark that this formulation is reminiscent of the measure for P- or CP-divisibility \cite{rivas2010entanglement} such that it is a sole function on the dynamical map so no optimization over state is required and the eigenvalues of the map obey a monotonic relation. 

Let us now discuss some of the properties of the above measure. First, we have that $(S^{\Lambda_t})^T = S^{\Lambda_t^\dagger}$. From here it is easy to see that the equality in Eq. \eqref{eq: non-markovianity witness} is saturated if the channel is unitary, which reduces to a orthogonality relation. This is obvious from the basic property of unitary matrices: $U^\dagger = U^{-1}$. Interestingly, it has been shown before that an axiom based on the orthogonality of the quasibistochastic map enables the reconstruction of the qubit and its dynamics \cite{onggadinata2023qubits}.

Second, an equivalent witness can be obtained by swapping the matrices in Eq. \eqref{eq: non-markovianity witness}, i.e., $S^{\Lambda_t}(S^{\Lambda_t})^T \leq \one$. This is clear as the eigenvalue of $AB$ is equal to $BA$. Moreover, both instances of the product matrix yield a symmetric matrix. The observation of the symmetry of the witness has a foundational implication that will be discussed later. 

Finally, we show that our measure is invariant under different choices of a minimal frame representation. Consider two minimal QPRs given by the pairs $(\{F^{(1)}_j\}, \{G^{(1)}_k\})$ and $(\{F^{(2)}_j\}, \{G_k^{(2)}\})$, where we use superscripts to denote the frame and dual frame belonging to the two different QPRs. The two representations can be shown to be connected by a pair of orthogonal quasibistochastic matrices $M$ and $\tilde{M}$ through $F^{(2)}_j = \sum_{k}M_{jk}F^{(1)}_k$ and $G^{(2)}_j = \sum_{k}\tilde{M}_{jk}G^{(1)}_k$. The properties of $M$ and $\tilde{M}$ can be deduced from ensuring that both representations obey the basic properties described in the preceding section. From here it is easy to show that the two QPRs of $\Lambda$ are connected via $S^{\Lambda,(2)} = MS^{\Lambda,(1)}\tilde{M}^T$, where $S^{\Lambda,(i)}_{jk} = \tr{F^{(i)}_j\Lambda[G^{(i)}_k]}$, $i=1,2$. As a result, we find that $(S^{\Lambda,(2)})^TS^{\Lambda,(2)}=(MS^{\Lambda,(1)}\tilde{M}^T)^T(MS^{\Lambda,(1)}\tilde{M}^T) = \tilde{M}(S^{\Lambda,(1)})^TS^{\Lambda,(1)}\tilde{M}^T$. Since this is simply an orthogonal transformation of a matrix by $\tilde{M}$, it does not change the matrix norm in Eq. \eqref{eq: non-markovianity measure} \cite{marshall1979inequalities}. We note that for a nonminimal representation, this may no longer hold, and the measure will be dependent on the choice of frames. Moreover, the nonuniqueness of the frame and dual frame pair can introduce further complications. Nevertheless, we stress that most studies involving the QPR are conducted with minimal frames.

\section{Examples}

Next we will see our measure of non-Markovianity in action over several paradigmatic examples with qubit and qutrit systems. Notice that the measure above is derived without any reference for a specific frame representation. Thus, the result works in general for any choice of quasiprobability representation. In order to make a specific calculation, here we pick the discrete Wigner representation with Wootters' representation for the qubit case \cite{wootters1987wigner} and Gross's for the qudit case (odd prime dimension) \cite{gross2006hudson}. In this representation, they belong to the family of normal quasiprobability representations as their frame and dual frame operators are proportional to each other \cite{zhu2016quasiprobability}. In fact, they are related via $G_j = dF_j$ since the frame is minimal. Wootters' discrete Wigner representation has frame operators given by
\begin{equation}
F_j = \frac{1}{4}\left[\one + (-1)^{j_1} \sigma_z  + (-1)^{j_2} \sigma_x + (-1)^{j_1 + j_2}\sigma_y \right]\, ,
\end{equation}
where $j \equiv (j_1,j_2) \in \mathbb{Z}_2 \times \mathbb{Z}_2$ and $\sigma_x$, $\sigma_y$, and $\sigma_z$ are qubit Pauli operators. The construction for Gross's discrete Wigner representation is more involved and since we are only working with a qutrit system in the example below, we show the frame operators for $d=3$ in the Appendix. 

We ascertain that the results below are frame independent by repeating the calculation with another representation, namely, the symmetric informationally complete (SIC) POVM representation \cite{kiktenko2020probability}.

\subsection{Pure decoherence model}

Let us start with the prototypical model of a two-level system in a bosonic bath environment undergoing pure decoherence dynamics \cite{luo2012quantifying}. The evolution of the system's state is given by the master equation 
\begin{equation}
\mathcal{L}[\rho(t)] = \frac{d}{dt}\rho(t) = \gamma(t)\left[\sigma_z\rho(t)\sigma_z - \rho(t)\right]\, ,
\end{equation}
where the time-dependent decay rate $\gamma(t) = -\frac{1}{G(t)}\frac{d}{dt}G(t)$, with $G(t)$ the decoherence function. From this the dynamical map reads
\begin{equation}
\Lambda_t : \rho(0) = \begin{bmatrix}
    \rho_{00} & \rho_{01} \\ \rho_{10} & \rho_{11}
\end{bmatrix} \longrightarrow \rho(t) = \begin{bmatrix}
    \rho_{00} & G(t)\rho_{01} \\ G(t)\rho_{10} & \rho_{11}
\end{bmatrix}\, .
\end{equation}
Note that the map $\Lambda_t$ is completely positive and trace preserving for $G(t)\leq 1$.

To find the non-Markovianity criterion, we first find the quasistochastic representation of $\Lambda_t$ following the Wootters representation:
\begin{equation}
S^{\Lambda_t} = \frac{1}{2}\begin{bmatrix}
    1 + G & 1-G & 0 & 0 \\
    1-G & 1+G & 0 & 0 \\
    0 & 0 & 1+G & 1-G \\
    0 & 0 & 1-G & 1+G
\end{bmatrix}\, ,
\end{equation}
where $G\equiv G(t)$ for simplicity. Notice that here $S^{\Lambda_t}$ is self-transpose; hence $\Lambda_t$ is its self-dual, i.e., $\Lambda_t^\dagger = \Lambda_t$. More importantly, the quasibistochastic map is represented non-negatively (same as the SIC POVM representation), which allows us to use the main result in Eq. \eqref{eq: basic monotonicity}. In fact, all the examples in this paper have a map that is non-negative. Calculating the left hand side of Eq. \eqref{eq: non-markovianity witness}, we find its eigenvalues are given by $\{1,1,G^2, G^2\}$. Using Eq. \eqref{eq: non-markovianity measure}, we obtain that
\begin{equation}
    \zeta(t) = 4G(t)\frac{d}{dt}G(t)  = -4\gamma(t)G^2(t)\, . 
\end{equation}
Therefore, it is clear that the Markovian condition is violated when $\gamma(t) < 0$. This criterion coincides with the BLP measure and in fact with the P- and CP-divisibility measures \cite{rivas2010entanglement}. In contrast, the non-Markovianity measures are different in general between the measure introduced here and the BLP measure, which read $\mathcal{N} = -4\int_{\gamma<0} dt\, \gamma(t)G^2(t)$ and $\mathcal{N}_{\text{BLP}} = -2\int_{\gamma <0} dt\, \gamma(t)G(t)$, respectively.

\subsection{Dissipation model}

Next we consider the same system and bath environment but with a system-environment interaction that allows for dissipation in the populations too. This evolution is governed by the time-local master equation 
\begin{eqnarray}
\frac{d}{dt}\rho(t) = & & -\frac{i}{2}s(t)[\sigma_+ \sigma_-,\rho(t)] \nonumber \\ & & \quad + \gamma(t)\left[\sigma_-\rho(t)\sigma_+ - \frac{1}{2}\big\{\sigma_+\sigma_-,\rho(t)\big\}\right]\, ,
\end{eqnarray}
where $\sigma_{\pm} = (\sigma_x \pm i\sigma_y)/2$ is the raising and lowering operators, $s(t) = -2\text{Im}[\dot{G}(t)/G(t)]$ is the time-dependent Lamb shift, and $\gamma(t) = -2\text{Re}[\dot{G}(t)/G(t)] = -\frac{2}{|G(t)|}\frac{d}{dt}|G(t)|$ is the decay rate. The dynamical map can be shown to be
\begin{equation}
    \Lambda_t:\rho(0) \rightarrow \rho(t) = \begin{bmatrix}
    \rho_{00} + [1-|G(t)|^2]\rho_{11} & G(t)^*\rho_{01} \\ G(t)\rho_{10} & |G(t)|^2\rho_{11}
\end{bmatrix}\, .
\end{equation}

Following the same calculation steps as the previous example, we find the eigenvalues of $(S^{\Lambda_t})^TS^{\Lambda_t}$ as $\{|G|^2, |G|^2, 1-|G|^2+|G|^4-\sqrt{(-1+|G|^2)^2(1+|G|^4)}, 1-|G|^2+|G|^4+\sqrt{(-1+|G|^2)^2(1+|G|^4)}\}$. From here, we find 
\begin{equation}
\zeta(t) = 8|G(t)|^3 \frac{d}{dt}|G(t)| = -4\gamma(t) |G(t)|^4
\end{equation}
and identify that non-Markovianity is observed when $\gamma(t)<0$. Again, the witness here coincides with the BLP and CP-divisibility measures \cite{breuer2016colloquium}.

\subsection{Random unitary channel}

In the last example, we look into evolution under a random unitary channel of a qubit and a qutrit system. The general qudit system has been studied in \cite{chruscinski2015non}. Here we consider the master equation
\begin{equation}
\frac{d}{dt}\rho(t) = \sum_{\alpha=1}^{d^2-1}\gamma_\alpha(t)\left[U_\alpha \rho(t) U_\alpha^\dagger - \rho(t)\right]\, ,
\end{equation}
where $U_\alpha \equiv U_{k,l} = \sum_{m=0}^{d-1}\omega^{mk}|m\rangle\langle m+k|$ is the Weyl (generalized spin) operator with $\alpha \equiv (k,l)$ via $\alpha = kd+l$, $k,l\in \mathbb{Z}_d$, and $\omega = e^{i2\pi/d}$. The dynamical map takes the form
\begin{equation}
    \Lambda_t[\rho] = \sum_{\alpha=0}^{d^2-1}p_\alpha(t) U_\alpha \rho U_\alpha^\dagger\, ,
\end{equation}
where 
\begin{eqnarray}
p_\alpha(t) &=& \frac{1}{d^2}\sum_{\beta=0}^{d^2-1} H_{\alpha\beta}\lambda_\beta(t)\, , \nonumber \\
\lambda_\beta(t) &=& \exp\left(\sum_{k=1}^{d^2-1}H_{\beta k}\Gamma_k(t)\right)\, ,
\end{eqnarray}
with $\Gamma_k(t) = \int_0^t \gamma_k(\tau) d\tau$ and $H_{\alpha\beta}$ a complex Hadamard matrix $H_{\alpha\beta}= \omega^{\alpha\times \beta}$, $\alpha\times \beta \equiv (k,l) \times (m,n) = kn - lm$. Note that for $p_\alpha(t)$ to be a probability distribution at all times and normalized, $\sum_{\alpha=0}^{d^2-1}p_\alpha(t) = 1$, we set $\lambda_0(t) = 1$. From the above relations, we also know that there are $d^2-1$ independent rates $\gamma_k(t)$, and $\gamma_0$ can be determined by the other $\gamma_k$'s via $\gamma_0 = -\sum_{k=1}^{d^2-1}\gamma_k$.

For $d=2$, the Weyl operators reduce to the identity and Pauli operators. In this case, we find the criteria for the Markovian condition to be
\begin{eqnarray}
\gamma_1(t) + \gamma_2(t) & \geq & 0 \, ,\nonumber \\
\gamma_1(t) + \gamma_3(t) & \geq & 0 \, ,\nonumber \\
\gamma_2(t) + \gamma_3(t) & \geq & 0 \, .
\end{eqnarray}
This coincides with the BLP and P-divisibility conditions, as expected \cite{chruscinski2013non}, but differs from the CP-divisibility condition, which is $\gamma_k(t) \geq 0$. This is an instance of the departure of the Markovian condition between information backflow and divisibility, with the former being weaker. 

In $d=3$, we find the Markovian criteria as
\begin{eqnarray}\label{eq: condition qutrit random unitary channel}
\gamma_1(t) + \gamma_2(t) + \gamma_4(t) + \gamma_5(t) + \gamma_7(t) + \gamma_8(t) & \geq & 0\, , \nonumber \\
\gamma_1(t) + \gamma_2(t) + \gamma_3(t) + \gamma_4(t) + \gamma_6(t) + \gamma_8(t) & \geq & 0\, , \nonumber \\
\gamma_1(t) + \gamma_2(t) + \gamma_3(t) + \gamma_5(t) + \gamma_6(t) + \gamma_7(t) & \geq & 0\, , \nonumber \\
\gamma_3(t) + \gamma_4(t) + \gamma_5(t) + \gamma_6(t) + \gamma_7(t) + \gamma_8(t) & \geq & 0\, .
\end{eqnarray}
To our knowledge, characterization based on BLP and P-divisibility appears to be highly nontrivial and only a sufficient (but not necessary) condition is obtained,
\begin{equation}
    \gamma_{i_1}(t) + \gamma_{i_2}(t) + \gamma_{i_3}(t) \geq 0
\end{equation}
for all triples $\{i_1,i_2,i_3\}\subset \{1,\dots,8\}$. Similar to the qubit case, the evolution is CP-divisible if all rates are non-negative $\gamma_\alpha(t)\geq 0$. 

\section{Discussions}

The monotonic relation in Eq. \eqref{eq: non-markovianity witness} provides an interesting perspective on non-Markovianity between classical and quantum theory. In the classical case, evolution is generally described by a stochastic map. A bistochastic map, a subset of all stochastic maps, has a special property such that its transpose returns a valid map as well. One can think of the transpose as an attempt to ``reverse" the transition probabilities between the input and output. Recent works \cite{coecke2017picturing, coecke2017time, chiribella2021symmetries} have also pointed out that the Hermitian adjoint and transpose of a quantum and a classical process, respectively, correspond to the time-reversal of the process. Interestingly, under the QPR the time-reversal notion has been unified for both theories with the transpose operation. Reversal of the process, however, does not always help in reversing the state of the system. This is also reflected in the monotonic relation, in both forms of $S^TS$ and $SS^T$, that will mix any probability distribution further. In contrast, applying forward and reverse quantum processes successively, or vice versa, on any state may result in a momentary revival of quantum features, albeit still leading to noise eventually.

It should be noted that reversal is not equivalent to the inverse, which is generally not a valid map except for unitaries.\footnote{For classical theory, unitary processes are described by permutation matrices, i.e., matrices with a single $1$ in each row and column and $0$ elsewhere.} Nevertheless, for an irreversible process one can still construct a map that approximately inverts the forward process known as Bayesian inference. Remarkably, the canonical quantum Bayesian inference, the Petz map, also contains $S^T$, which might imply that it is a useful ingredient in retrieving lost information \cite{cenxin2023quantum}.

In formulating the condition for Markovianity, it is also required that quantum maps are represented non-negatively. This is seemingly counterintuitive to current perception, which widely believes that the appearance of negativity is necessary for many quantum phenomena. Thus, one would also think that quantum non-Markovianity, a nonclassical phenomenon, would occur due to negativity in the dynamical map. It would be interesting to show if a negative quasi-stochastic map corresponding to a Markovian evolution would still satisfy the monotonic relation in Eq. \eqref{eq: basic monotonicity}. Alternatively, it would also be interesting if it is always possible to find a frame representation that will return a non-negative quasistochastic matrix. 

Although the QPR of the dynamical maps studied above are represented non-negatively, we would like to point out that the generator $\mathcal{L}$ has a nonclassical representation. More explicitly, we find $L=[L_{jk}]$, $L_{jk} = \tr{F_j\mathcal{L}[G_k]} \in \mathbb{R}$, which is a $d^2\times d^2$ matrix that satisfies $\sum_{j}L_{jk} = 0\, \forall\,k$. This is the quantum analog of the classical Kolmogorov generator \cite{van1992stochastic} that additionally satisfies $L_{jk} \geq 0$ for $j\neq k$, i.e., its off-diagonal elements are non-negative. For the pure decoherence and random unitary channel models, we observe that the off-diagonal entries of $L$ become negative when the non-Markovianity condition is satisfied and non-negative otherwise. However, for the dissipative model the matrix $L$ has negative nondiagonal entries in general. This suggests that negativity in the generator appears to be necessary but not sufficient for non-Markovianity. This observation has also been pointed out in \cite{benatti2024quantum}, which studied its connection to classical and quantum P-divisibility. It remains unclear how exactly negativity in the generator of quantum evolution contributes to the degree of non-Markovianity. We believe that the study using the QPR would be crucial in gaining a fundamental understanding of the difference between classical and quantum processes. 


\section{Conclusions}

In this work, we have proposed a new witness and measure of non-Markovianity using the formalism of quasiprobability representation. In particular, the measure detects information backflow through the breaking of the monotonic relation of the product of a quasibistochastic map, representing the dynamical map, and its transpose. The measure is explicitly dependent only on the dynamical map, thus forgoing any optimization task that plagues much of the previously proposed measure. This is particularly useful when studying the evolution of a quantum system of large dimensions. We illustrated the use over several paradigmatic examples with the explicit calculation done in the discrete Wigner and SIC POVM representation, the two most widely used frame representations for discrete quantum mechanics. We further discussed on the nonclassicality of the evolution and believe that the process generator, the one possessing the negativity, should be a focus for future work in its role in quantum non-Markovianity.

It should be noted that our study of non-Markovian dynamics is to the extent of the information backflow notion, and there are several other concepts of non-Markovianity that are not explored here. Although the concept of divisibility seems to be the popular definition of Markovianity, there remains a large debate to this day on the proper definition of Markovianity \cite{li2018concepts}. We also would like to mention that concepts beyond quantum dynamical maps, used primarily in divisibility and information backflow approach, also exist, e.g., the quantum regression formula \cite{chruscinski2022dynamical,lonigro2022quantum}. In this case, we believe that recent advances in other forms of quasiprobability, namely, Kirkwood-Dirac quasiprobability \cite{arvidsson2024properties}, can be an indispensable tool to study quantum regression formula as it concerns time-ordered correlation functions, a quantity that can be efficiently described using Kirkwood-Dirac quasiprobability \cite{wang2024snapshotting}. We leave this for future work.

\hfill

\section*{Acknowledgements}

We thank Dagomir Kaszlikowski and Lim Zi Yao for discussions. This work was supported by the Ministry of Education, Singapore, under its Academic Research Fund Programme (Grant No. RG154/24). K.O. was supported by grants from the National Research Foundation, Singapore and A*STAR under its Quantum Engineering Programme 2.0 (Grant No. NRF2021-QEP2-02-P07) and the Singapore Ministry of Education Academic Research Fund (Grant No. MOE-T2EP50222-0017).


\begin{appendix}
\section{Appendix: Discrete Wigner representation for qutrit}

The following are the frame operators for the discrete Wigner representation in $d=3$:
\begin{eqnarray}
& F_{00} = \frac{1}{3}\begin{bmatrix} 1 & 0 & 0 \\ 0 & 0 & 1 \\ 0 & 1 & 0\end{bmatrix}\, , \quad  & F_{01} = \frac{1}{3}\begin{bmatrix}0 & 0 & 1 \\ 0 & 1 & 0 \\ 1 & 0 & 0 \end{bmatrix}\, , \nonumber \\
& F_{10} =  \frac{1}{3}\begin{bmatrix}0 & 1 & 0 \\ 1 & 0 & 0 \\ 0 & 0 & 1 \end{bmatrix}\, , \quad & F_{10} = \frac{1}{3}\begin{bmatrix} 1 & 0 & 0 \\ 0 & 0 & -\omega \\ 0 & \omega^2 & 0\end{bmatrix}\, , \nonumber \\
& F_{11} = \frac{1}{3}\begin{bmatrix}0 & 0 & \omega^2 \\ 0 & 1 & 0 \\ -\omega & 0 & 0 \end{bmatrix}\, , \quad & F_{12} = \frac{1}{3}\begin{bmatrix}0 & -\omega & 0 \\ \omega^2 & 0 & 0 \\ 0 & 0 & 1 \end{bmatrix}\, , \nonumber \\
& F_{20} = \frac{1}{3}\begin{bmatrix} 1 & 0 & 0 \\ 0 & 0 & \omega^2 \\ 0 & -\omega & 0\end{bmatrix}\, , \quad & F_{21} = \frac{1}{3}\begin{bmatrix}0 & 0 & -\omega \\ 0 & 1 & 0 \\ \omega^2 & 0 & 0 \end{bmatrix}\, , \nonumber \\
& F_{22} =\frac{1}{3}\begin{bmatrix}0 & \omega^2 & 0 \\ -\omega & 0 & 0 \\ 0 & 0 & 1 \end{bmatrix}\, , \quad & 
\end{eqnarray}
where $\omega = e^{i2\pi/3}$ is the cube root of unity.

\end{appendix}


\bibliography{ref.bib}


\end{document}